\documentclass[journal]{IEEEtran}

\usepackage{cite}
\usepackage{graphicx}
\usepackage{hyperref}
\usepackage{amsmath}
\usepackage{amssymb}
\usepackage{array}
\usepackage{url}
\usepackage{color}
\usepackage{algpseudocode}

\begin{document}
\title{Sidelobe Modification for an Offset Gregorian Reflector System using a Reconfigurable Intelligent Surface-Equipped Subreflector}
\author{
S.W.~Ellingson and
A.J.~Yip\thanks{The authors are with the Dept.\ of Electrical and Computer Engineering, Virginia Tech, Blacksburg, VA, 24061 USA, e-mail: ellingson@vt.edu.}
}


\maketitle


\begin{abstract}
In past work, we described the use of a reconfigurable intelligent surface (RIS) mounted on the rim of an axisymmetric prime focus-fed reflector to create nulls in the close-in sidelobes.
In this paper, we show that similar performance is possible in an offset Gregorian reflector system using a RIS on the rim of the subreflector.
Applications include radio astronomy, where offset Gregorian reflectors are common and observations are subject to deleterious levels of interference from satellites entering through sidelobes.  
We show that an efficient RIS replacing the outer one-third of the subreflector surface, employing passive elements with 1-bit phase-only control, can create a null in the peak of the second sidelobe in the quiescent pattern.
This is achieved using a simple unconstrained optimization algorithm to set the states of the RIS elements.
The algorithm yields a deep null with just 0.2~dB reduction in main lobe directivity, despite lacking any constraints on main lobe pattern. 
Compared to our previous approach of mounting the RIS on the rim of the main reflector, the subreflector-based approach demonstrated in this paper requires a much smaller RIS and can implemented in existing systems by replacing the subreflector.   
\end{abstract}


%

\section{Introduction}
\label{sIntro}

A reconfigurable intelligent surface (RIS) is an electromagnetic aperture whose scattering can be dynamically controlled using electronics.
An important class of these devices includes reconfigurable reflectarrays; i.e., passive arrays comprised of sub-wavelength unit cells with electronically-variable terminations; see e.g. \cite{Hum+1_1401} and references therein.
In a previous paper \cite{ES21}, we showed that a reflectarray-type RIS along the rim of a circular axisymmetric prime focus-fed reflector antenna system can be used to create an electronically-steerable null in the combined pattern. 
An application of this technique is mitigation of interference from satellites to radio astronomy. 
In subsequent work, 
we demonstrated that the concept can be implemented in practical hardware and exhibits good performance in full-wave electromagnetic simulations \cite{B+24,S+25}; 
we developed and demonstrated practical algorithms for setting RIS element states for driving deep nulls simultaneously with main lobe constraints \cite{LBE24,LBE25}; and 
we demonstrated that this concept can be applied to existing reflectors using rim-mounted flat-panel rectangular RISs \cite{Yip24}.

\begin{figure}
\begin{center}
\begin{tabular}{cc}
\includegraphics[width=0.42\columnwidth]{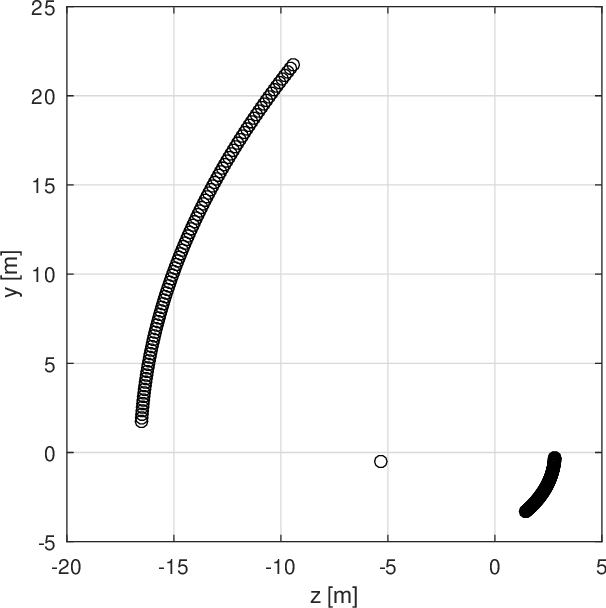} & 
~~ \includegraphics[width=0.42\columnwidth]{fig/xy.epsc} 
\end{tabular}
\end{center}
\caption{
System considered in this paper.
Boresight is in the $+z$ direction.
\emph{Left:} Side view (cross-section in the $x=0$ plane). The common focus is at the global origin, and the feed location is the circle marker near $(y,z)=(0,-5)$~m. 
\emph{Right:}  Front view (projection into the $z=0$ plane). The rim of the main reflector is in black.  The original (non-reconfigurable) and RIS portions of the subreflector are in red and blue, respectively. 
\label{fConcept}
}
\end{figure}

In this paper, we consider a similar technique, now applied to dual reflector systems.
In particular, we consider an offset Gregorian system in which the RIS is applied to the rim of the subreflector, without modification to the main reflector.
The concept is illustrated in Figure~\ref{fConcept} (see also Figure~\ref{fsxy}).
This work is motivated by two principal considerations.
First, many modern radio telescopes employ offset Gregorian optics, owing to the attractive combination of unblocked aperture, sky-directed spillover, and relatively compact size \cite{E16}; this includes 
the Green Bank Telescope \cite{GBT}, 
the Allen Telescope Array \cite{ATA},
MeerKAT \cite{DJ12}, 
and new telescopes in development including 
the Next Generation VLA \cite{ngVLA}.\footnote{The general strategy of a RIS along the rim of the subreflector is also applicable to other dual reflector schemes including the Cassegrain optics employed by the Expanded Very Large Array (VLA) \cite{EVLA}, but is not explicitly considered in this paper.}
The second consideration is cost and efficacy of installation on existing systems.  Since all useful power intercepted by the main reflector is also intercepted by the subreflector, and since the subreflector is much smaller, effective nulling can be achieved using a smaller RIS with fewer reconfigurable elements, and can be implemented by replacing or modifying the subreflector.  
 
\section{Reference System}
\label{sRS}

To demonstrate the concept, we employ the offset Gregorian system described in Figure~2, Figure~7, and Table~I of \cite{BP94}.
Although not strictly necessary, it is useful to set a nominal frequency of operation, so that dimensions can be expressed in meters; here, we choose 1.5~GHz (wavelength $\lambda=0.2$~m).
Front and side views of this system are shown in Figure~\ref{fConcept}.
Using the variables employed in \cite{BP94}, 
the paraboloidal main reflector of this system is circular (in projection) with 
diameter $D=20$~m ($100\lambda$), 
focal length $F=16.56$~m ($82.8\lambda$), and 
center offset $d_0=11.74$~m ($58.7\lambda$).
The relevant subreflector parameters are
eccentricity $e=0.49$,
half-distance between focii $c=2.678$~m ($13.39\lambda$), 
rim half-angle $\theta_e=11.95^{\circ}$, and
orientation angles
$\alpha=-15.87^{\circ}$ and
$\beta=+5.4^{\circ}$.
The feed is a Huygens source, pointed at the center of the subreflector, polarized in the vertical ($x=0$) plane, with electric field multiplied by $\exp{\left[kb\left(\cos\theta_f-1\right)\right]}$ where $k=2\pi/\lambda$, $\theta_f$ is the angle measured from direction of maximum feed directivity, and $b=1.661$~m, yielding 10~dB taper at the subreflector rim; i.e., at $\theta_f=\theta_e$. 

As shown in Figure~9 of \cite{BP94}, and confirmed in Section~\ref{sMOA} (Figure~\ref{fEx1}), this system achieves 
directivity of about 48.5~dBi 
and an $H$-plane copolarized pattern with 
half-power beamwidth (HPBW) of about $0.7^{\circ}$. 

In Section~\ref{sResults}, this system is modified to model the addition of a RIS along the subreflector rim.

\section{Method of Analysis}
\label{sMOA}

The directivity, main lobe shape, and characteristics of the close-in sidelobes of an electrically-large reflector antenna system can be accurately determined using physical optics (PO; see e.g.\ \cite{ST13}).
The reference system is analyzed using transmit-mode iterative PO analysis software developed by the authors.  This software first computes PO equivalent currents on the subreflector due to the feed.  Next, the PO equivalent currents on the main reflector -- i.e., those due to the subreflector equivalent currents -- are computed.  Finally, the far-field electric field is computed by integration over the main reflector currents.  The gridding for both subreflector and main reflectors consists of approximately square flat patches with side length approximately $0.2\lambda$. 

\begin{figure}
\centerline{\includegraphics[width=0.8\columnwidth]{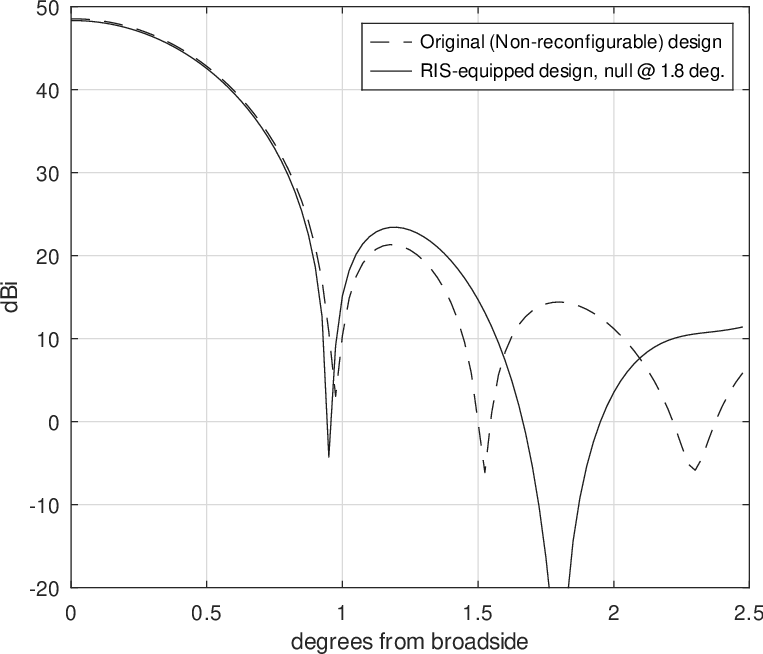}}
\caption{
$H$-plane copolarized patterns of (a) the original (reference) system and (b) the RIS-equipped system driving a null at $1.8^{\circ}$ using the algorithm defined in Section~\ref{sSSA}.
\label{fEx1}
}
\end{figure}

Figure~\ref{fEx1} shows the $H$-plane pattern of the reference system at 1.5~GHz, limiting view to the first few sidelobes around the main lobe where the PO approximation is known to be accurate.  
The computed gain, beamwidth, sidelobe levels and overall pattern characteristics are all in close agreement to the results shown in \cite{BP94}, which were also obtained by iterative PO.  

The RIS along the rim of the subreflector is modeled using the same technique employed in \cite{ES21} and subsequently validated in \cite{B+24,S+25} and references therein.
Specifically, the relevant portion of the subreflector surface is regridded with patches having sidelength $0.5\lambda$ as opposed to $0.2\lambda$.  
As in \cite{ES21}, the motivation for $0.5\lambda$ quantization of the reconfigurable portion of the surface is simply that technologies that might be used to implement reconfigurability would normally consist of unit cells having approximately this periodicity in order to satisfy the Nyquist condition for full sampling of the available aperture.
A coefficient $c_n$ is applied to the field scattered by the $n$-th element, with $n=1, 2,  ... N$ where $N$ is the number of reconfigurable elements.  Thus, $c_n=\pm 1$ for a perfectly-efficient reconfigurable element with 1-bit phase-only control.  Using this approach, analysis of the RIS-equipped system differs from the reference system only in gridding of the reconfigurable portion of the subreflector and application of the $c_n$'s to the associated patches (corresponding to RIS elements) in the PO integral.  

\section{State-Setting Algorithm}
\label{sSSA}

To implement sidelobe modification, an algorithm is required to determine the $c_n$'s.  
For the main reflector RIS implementation of \cite{ES21}, a simple ``serial search'' algorithm was sufficient to drive a deep null at a specified position.
Subsequent work resulted in algorithms capable of multiple constraints; e.g., multiple nulls with or without constraints on main lobe characteristics
\cite{LBE24,LBE25}. 
Unfortunately, these approaches are not directly applicable to systems in which the RIS is implemented on a subreflector.
This is for two reasons:
First, each element of the subreflector RIS illuminates the entire main reflector, so the contribution of each element to a particular angle in the far field pattern can be determined only indirectly through integration over the entire surface of the main reflector.  
Second, each element of the subreflector RIS intercepts a relatively large fraction of the available power, making it more difficult to find combinations of elements whose sum contributions precisely cancel the contribution of the non-reconfigurable portion of the subreflector in the far field, especially for 1-bit phase-only control of elements.

The following alternative approach has been found to be effective for 1-bit phase-only control, and is used to obtain the results reported in this paper.
To simplify the discussion, let us limit our attention to the far-field $H$-plane copolarized pattern in transmit-mode operation.
In order to create a null in a specified direction $\psi_0$ in the pattern, one first calculates ${\bf E}_0(\psi_0)$, the contribution to the copolarized far field electric field due to the entire non-reconfigurable portion of the subreflector, and also ${\bf E}_n(\psi_0)$ for $n=1, 2, ... N$, the contribution to the copolarized far field electric field due to the RIS element with index $n$ and with $c_n=+1$.
The ${\bf E}_n(\psi_0)$'s are then ranked by magnitude.
Let $p(m)$ be this ranking, such that 
$p(1)$ is the value of $n$ for which $\left|{\bf E}_n(\psi_0)\right|$ is largest, 
$p(2)$ is the value of $n$ for which $\left|{\bf E}_n(\psi_0)\right|$ is second largest, and so on, so that 
$p(N)$ is the value of $n$ for which $\left|{\bf E}_n(\psi_0)\right|$ is smallest.
The algorithm is then as follows:
\pagebreak
\begin{algorithmic}
\State ${\bf E}(\psi_0) \leftarrow {\bf E}_0(\psi_0)$
\For{$m=1$ to $N$}
  \State $c_{p(m)} \leftarrow +1$
  \If{$\left|{\bf E}(\psi_0)-{\bf E}_{p(m)}(\psi_0)\right|<\left|{\bf E}(\psi_0)\right|$} 
    \State $c_{p(m)} \leftarrow -1$
    \EndIf
  \State ${\bf E}(\psi_0) \leftarrow {\bf E}(\psi_0) + c_{p(m)} {\bf E}_{p(m)}(\psi_0)$
\EndFor
\end{algorithmic}
In other words, the pre-computed element contributions are considered in order from largest magnitude to smallest magnitude, and $c_n$ is set to the quiescent value of $+1$ unless setting $c_n$ to $-1$ reduces the present sum magnitude of the electric field.  The key feature of this algorithm is following the sequence in order of decreasing element contribution magnitude, as this prevents the solution from simply ``jittering'' around a relatively weak null.  A pleasant side benefit of this algorithm is that a relatively small number of $c_n$'s will be changed (i.e., set to $-1$), so that there is relatively little perturbation in the main lobe; this is demonstrated in Section~\ref{sResults}.

Note that this algorithm is presented merely as a simple way to demonstrate the technical concept, and we anticipate that improved algorithms, including those with main lobe constraints analogous to those that we have developed for focus-fed axisymmetric implementations \cite{LBE24,LBE25}, can be developed. 

\section{Demonstration of the RIS-Equipped Subreflector}
\label{sResults}

\begin{figure}
\centerline{\includegraphics[width=0.7\columnwidth]{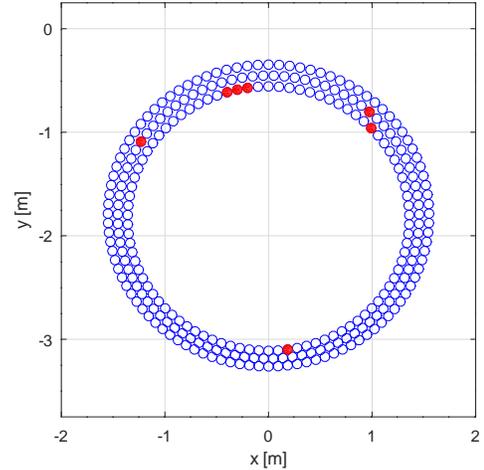}}
\caption{
Front view (projection into the $z=0$ plane) of the subreflector. Circle markers indicate the locations of the reconfigurable elements comprising the RIS.  
The red-filled markers indicate the elements for which $c_n=-1$ in order to drive the null at $1.8^{\circ}$ as shown in Figure~\ref{fEx1}.
\label{fsxy}
}
\end{figure}

We now consider the performance of the reference system with a RIS-equipped subreflector.  For this study, the radial width of the RIS was arbitrarily set to 0.3~m, as shown in Figure~\ref{fConcept} (also shown in Figure~\ref{fsxy}).  This results in a RIS model consisting of $N=282$ elements, each having dimensions approximately $0.5\lambda \times 0.5\lambda$. As an initial check, the performance of the RIS-equipped system was computed for $c_n=+1$ for all $n$, yielding (as expected) negligible difference from the result shown for the original reference system in Figure~\ref{fEx1}.

The analysis of Section~\ref{sMOA} is now repeated using a set of $c_n$'s determined using the algorithm of Section~\ref{sSSA} to drive a null at $1.8^{\circ}$. 
This corresponds to the peak of the second sidelobe in the pattern of the reference system.
The result is shown in Figure~\ref{fEx1}.
Note that the desired null has been created, and has been done so with little perturbation of the main lobe.
The main lobe directivity has been reduced by just 0.2~dB relative to that of the reference system.
Presumably this can be further improved using a state-setting algorithm that incorporates main lobe constraints.
Just 7 of 282 element states were changed (i.e., $c_n$ set to $-1$) by the state-setting algorithm to achieve this performance. 
Figure~\ref{fsxy} shows the location of these elements. 

\section{Conclusions}

We have demonstrated the ability of a RIS-equipped subreflector to drive a deep null at the peak of the second sidelobe of the quiescent pattern of large offset Gregorian reflector system, with little perturbation of the main lobe. In this example (20~m main reflector with 3.2~m subreflector at 1.5~GHz), the RIS consisted of just 282 reconfigurable elements -- orders of magnitude less than required for the main reflector RIS implementation considered in \cite{ES21}.  This capability could be implemented in existing systems by replacing the subreflector or by installing flat panel RIS ``outriggers'' as described in \cite{Yip24}.

\section*{Acknowledgment}

This work was supported in part by the U.S. National Science Foundation under Grant AST-2128506.



\end{document}